\documentclass[useAMS,usenatbib]{mn2e}

\usepackage{times}
\usepackage{graphicx}
\usepackage{txfonts}
\usepackage{natbib}
\newcommand{\tssp}{$t_\mathrm{SSP}$}
\newcommand{\ty}{$t_\mathrm{y}$}
\newcommand{\zssp}{$[Z/H]_\mathrm{SSP}$}
\newcommand{\tc}{$T_\mathrm{c}$}
\newcommand{\hb}{H$\beta$}
\newcommand{\mgfe}{[MgFe]}
\newcommand{\hi}{{\sc H\,i}}
\newcommand{\mhi}{$M$(\hi)}
\newcommand{\mhilb}{$M$(\hi)/$L_B$}
\newcommand{\msun}{M$_\odot$}
\newcommand{\afe}{[$\alpha$/Fe]}
\newcommand{\my}{$M_\mathrm{y}/M_\mathrm{t}$}
\title[Cold gas and young stars in tidally-disturbed ellipticals at $z$=0]{Cold gas and young stars in tidally-disturbed ellipticals at $z$=0}
\author[P. Serra and T.~A. Oosterloo]{P. Serra\thanks{E-mail: serra@astron.nl} and T.~A. Oosterloo
\\
ASTRON, Postbus 2, 7990 AA Dwingeloo, the Netherlands}
\begin{document}

\date{Accepted 2009 October 13. Received 2009 October 13; in original form 2009 October 02}

\pagerange{\pageref{firstpage}--\pageref{lastpage}} \pubyear{2009}

\maketitle

\label{firstpage}

\begin{abstract}
We present an analysis of the neutral hydrogen and stellar populations of elliptical galaxies in the \cite{2009arXiv0908.1382T} sample. Our aim is to test their conclusion that the continuing assembly of these galaxies at $z$$\sim$0 is essentially gas-free and not accompanied by significant star formation. In order to do so, we make use of \hi\ data and line-strength indices available in the literature. We look for direct and indirect evidence of the presence of cold gas during the recent assembly of these objects and analyse its relation to galaxy morphological fine structure.

We find that $\geq$25\% of ellipticals contain \hi\ at the level of \mhi$>$10$^8$\msun, and that \mhi\ is of the order of a few percent of the total stellar mass. Available data are insufficient to establish whether galaxies with a disturbed stellar morphology are more likely to contain \hi. However, \hi\ interferometry reveals very disturbed gas morphology/kinematics in all but one of the detected systems, confirming the continuing assembly of many ellipticals but also showing that this is not necessarily gas-free. We also find that all very disturbed ellipticals have a single-stellar-population-equivalent age $<$4 Gyr. We interpret this as evidence that $\sim$0.5-5\% of their stellar mass is contained in a young population formed during the past $\sim$1 Gyr. Overall, a large fraction of ellipticals seem to have continued their assembly over the past few Gyr in the presence of a mass of cold gas of the order of 10\% of the galaxy stellar mass. This material is now observable as neutral hydrogen and young stars.
\end{abstract}


\section{Introduction}
The role of dissipationless galaxy merging in the assembly of early-type galaxies is currently under debate. On the one hand, it appears to be necessary to produce the boxy, slowly-rotating galaxies which populate the high-mass end of the red sequence \citep[e.g., ][and references therein]{1997AJ....114.1771F,2009ApJS..182..216K}. On the other hand, dissipative processes are required to explain a number of structural, kinematical and stellar-population properties of intermediate- and low-mass early-type galaxies \citep[e.g., ][and refernces therein]{1992ApJ...399..462B,2007MNRAS.379..401E,2009ApJS..181..135H}. Furthermore, the tight scaling relations observed in the local Universe place a rather conservative upper limit on the fraction of stellar mass assembled via dissipationless merging \citep[e.g., ][]{2003MNRAS.342..501N,2009arXiv0908.1621N}. Finally, cold gas is actually observed in a large fraction of early-type galaxies in the nearby Universe \citep[e.g., ][]{2006MNRAS.371..157M,2007MNRAS.377.1795C}.

Based on deep imaging of a volume-limited sample of nearby elliptical galaxies with M$_B \leq-20$, and following earlier work by, e.g., \cite{1983ApJ...274..534M} and \cite{1992AJ....104.1039S}, \citet[hereafter T09]{2009arXiv0908.1382T} find signatures of recent dynamically-violent assembly in a remarkably large fraction of objects, $\sim$75\%. Analysing galaxies' $B-V$ colour, they argue that ellipticals outside cluster environments continue to grow at $z$=0 mostly through gas-free accretions. If correct, this conclusion would have implications for a number of properties of local ellipticals. This follows from a large number of numerical studies showing that the presence of even a modest amount of gas ($\sim$10-20\% of the baryonic mass)  in a binary merger can change the stellar mass-distribution and orbits in the early-type remnant relative to the gas-free case \citep[e.g., ][]{2005MNRAS.360.1185J,2007MNRAS.376..997J,2009ApJS..181..135H}. Observable quantities such as ellipticity, isophote disciness/boxiness, anisotropy and light profile are affected by the gas content of the progenitor galaxies.

Although most of these numerical simulations focus on binary major or minor mergers, early-type galaxies in a $\Lambda$CDM Universe form following a sequence of many interactions with widely varying mass ratio and degree of dissipation. In other words, galaxies are much more likely to grow via continuous assembly of smaller units of varying gas content rather than a few major events, in a way strongly dependent on their environment \citep[e.g., ][]{2006MNRAS.366..499D,2007ApJ...658..710N,2009MNRAS.396.1972P}. Within this framework, as in the binary merger case, a galaxy formed following a gas-free path (e.g., one residing in the centre of a massive cluster, where little cold gas can penetrate) should be significantly different from one that experienced dissipation \citep[e.g.,][]{2007A&A...476.1179B}. It is therefore interesting to estimate the mass of cold gas present during the continuing assembly of ellipticals over the last few Gyr and test T09 conclusion that this is indeed negligible.

The approach taken in this Letter is to look for neutral hydrogen and young stellar populations in galaxies in the T09 sample. Both properties could be regarded as marginally important with respect to the broad cosmological picture of galaxy evolution. After all, work over the past decades shows that young stars and cold gas amount to just a few percent of the total galaxy stellar mass of $z$=0 ellipticals. However, as explained above, even modest amounts of gas can have important effects on fundamental properties of these objects. We would like to clarify what kind of data analysis is necessary in order to detect this gas. Work along this line has been carried out by a number of authors during the past few years. We summarise it together with our results in the discussion section.

\section{Neutral hydrogen}

   \begin{figure}
   \centering
   \includegraphics[width=9cm]{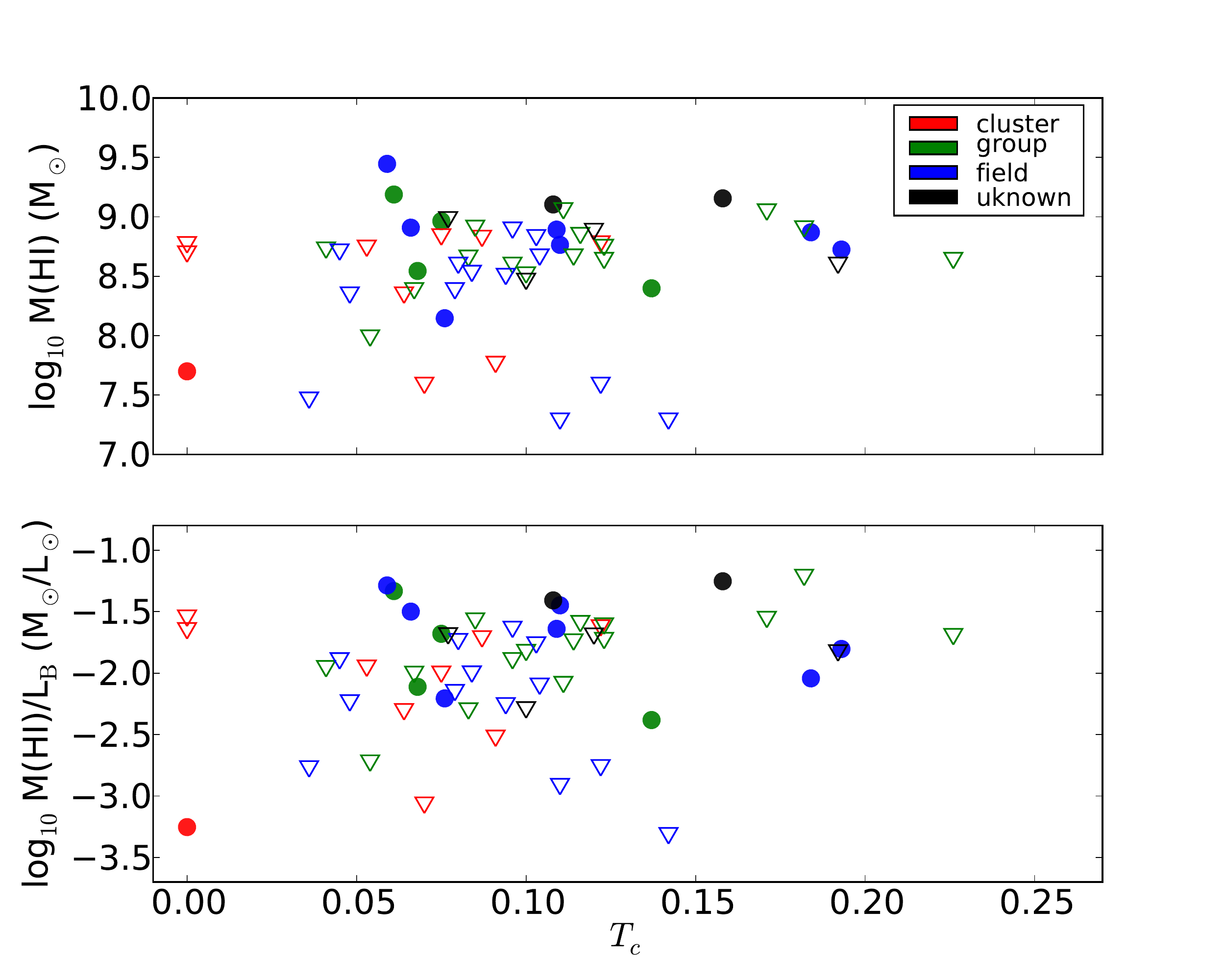}
      \caption{\hi\ content of galaxies in the T09 sample. Circles are \hi-detected galaxies, triangles are upper limits on the \hi\ mass. Colour represents the environment as reported by T09.}
         \label{fig1}
   \end{figure}

All galaxies in the T09 sample have been observed in the 21-cm line by the HIPASS survey \citep{2004MNRAS.350.1195M}. The noise in the HIPASS spectra is $\sim$15 mJy in $\sim$13-km/s-wide channels. It corresponds to a 3$\sigma$ flux of $\sim$2.4 Jy$\cdot$km/s over 200 km/s, which we take as the typical width of the \hi\ profile. Therefore, the HIPASS \mhi\ detection limit varies in the range 0.1-1.2$\times$10$^9$ \msun\ as a function of distance for the T09 sample\footnote{\mhi=$2.36\times10^5 F$(\hi)$/\mathrm{(Jy\cdot km/s)} \ (d/\mathrm{Mpc})^2$ \msun}. This is not very deep and indeed, after inspection of the HIPASS spectra, we find that only 2 galaxies are detected: IC~1459 and IC~4889. We searched the literature for deeper \hi\ observations. This resulted in 12 more detections and 9 upper limits deeper than those provided by HIPASS (derived with the same criterion described above). In one more galaxy, NGC~6868, \hi\ is detected in absorption against the central radio continuum source (Oosterloo, priv.~comm.) and estimating its \mhi\ is not possible with current data. We refer to Table \ref{tab1} for details on the \hi\ data.

In Fig.\ref{fig1} we plot \mhi\ and \mhilb\ against the tidal parameter \tc\ introduced by T09. \tc\ is designed as a way of quantifying the amount of morphological fine structure and is larger for galaxies with higher disturbance. In the figure, colour represents galaxy environment following the T09 division of the sample in cluster, group, field and unknown environment. Overall, we find \hi\ in/around $\sim$1/4 of ellipticals, and mostly outside the cluster environment. Because of the inhomogeneity of the data, we conclude that $\geq$25\% of ellipticals contain \hi\ at the 10$^8$-\msun\ level. \mhi\ can be as high as a few times 10$^9$ \msun, corresponding to at most a few percent of the total stellar mass. Given the many upper limits, we cannot establish whether highly-disturbed ellipticals host more \hi\ in/around their stellar body than more relaxed objects.

Most of the \hi\ data used here are taken with single-dish telescopes. These have a beam of several arcmin $FWHM$ and we have discarded 4 detections that are likely caused by the presence of a late-type neighbour within the same beam. However, a number of observations taken with interferometers are available and allow us to study the actual spatial distribution of the \hi\ around about half of the detected galaxies. Here we comment on individual galaxies for which radio interferometry is available.

IC~1459 shows a number of \hi\ filaments surrounding the stellar body; the galaxy resides in a rich group with many gas-rich disc galaxies. IC~4889 hosts a strongly-warped disc/ring, but the poor spatial resolution of the ATCA data may hide a more complicated configuration. NGC~2865 is a known recent-merger remnant with \hi\ associated to a number of stellar shells. NGC~2974 hosts a regular \hi\ ring aligned with the stellar body. NGC~4472, the brightest galaxy in the Virgo cluster, shows an \hi\ cloud that suggests on-going interaction with a dwarf neighbour. NGC~5018 shows a filament of gas stretching across its stellar body and connecting two neighbours. NGC~5077 is likely in the process of acquiring gas from a late-type companion as a broad tail of \hi\ connects the two systems and \hi\ clouds are scattered around them. NGC~5903 hosts kinematically-disturbed \hi\ distributed along the galaxy minor axis and two gas tails connected to the edge of this distribution.

Overall, interferometric data show a disturbed \hi\ configuration, signature of on-ongoing or recent ($\sim$1 Gyr) gas accretion, in all but one of the detected ellipticals. Considering gas consumption in star formation and tidal-tail gas lost to the inter-galactic medium, it is plausible that many ellipticals have continued their assembly until recently in the presence of a mass of cold gas as high as $\sim$10\% of their total stellar mass. Given the many high upper limits on \mhi, deeper \hi\ observations would be required to establish the actual fraction of ellipticals that host neutral hydrogen at this level in the T09 sample. However, we can state that this is $\geq$25\%, in agreement with deep observations where \hi\ is found in about half of the targets \citep[][]{2006MNRAS.371..157M}. It is clear that the presence of significant amounts of cold gas during the recent assembly of these systems is fairly common outside cluster environments.

\section{Stellar populations}

Fully corrected line-strength indices measured from optical spectra and brought onto the Lick/IDS system are available in the literature for 38/55 galaxies in the T09 sample. For 34 of them indices within an $R_e$/10 aperture are given by \cite{2005ApJ...621..673T} or \cite{2006A&A...445...79A}. Indices for a few additional objects and measured over slightly different apertures are available in \cite{2007MNRAS.377..759S} and \cite{2008A&A...483...57S}. We refer to Table \ref{tab1} for details on the line-strength data.

\begin{figure}
   \centering
   \includegraphics[width=9cm]{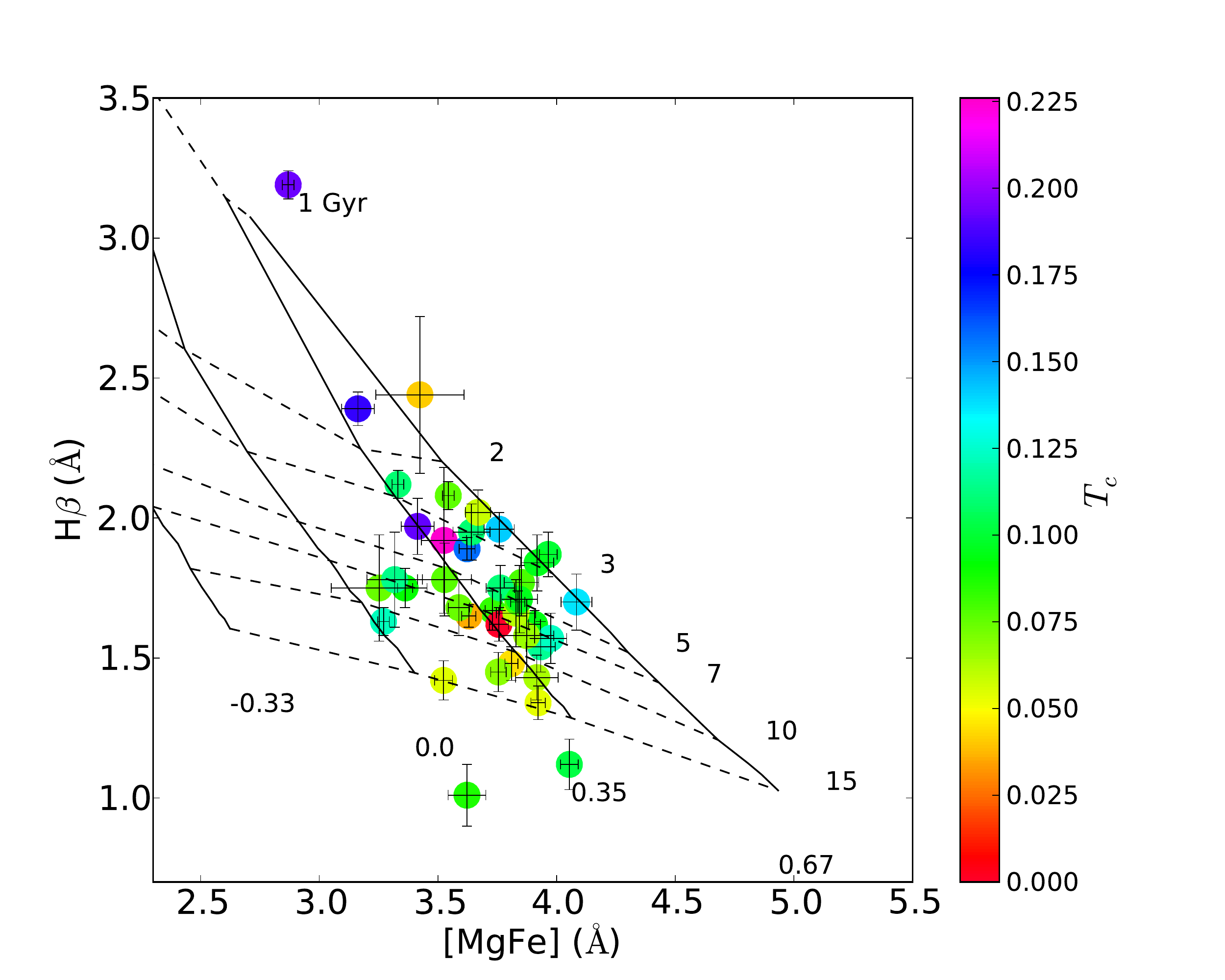}
%      \caption{\hb\ vs. \mgfe\ for the 38 galaxies in the T09 sample with available line-strength indices. Colour codes the tidal parameter \tc\ defined in T09. The model grid represents SSP models from \cite{2003MNRAS.339..897T}. Solid lines are models of fixed metallicity with $[Z/H]$ labelled at the bottom of the grid. Dashed lines are models of fixed stellar age with $t$ in Gyr labelled on the right of the grid.}
      \caption[short caption]{\hb\ vs. \mgfe\ for the 38 galaxies in the T09 sample with available line-strength indices. Colour codes the tidal parameter \tc\ defined in T09. The model grid represents SSP models from \cite{2003MNRAS.339..897T}. Solid lines are models of fixed metallicity with $[Z/H]$ labelled at the bottom of the grid. Dashed lines are models of fixed stellar age with.}
         \label{fig2}
   \end{figure}

In Fig.\ref{fig2} we plot \hb, an age-sensitive index, against \mgfe, a metallicity indicator nearly insensitive to \afe\ variations\footnote{\mgfe=$\sqrt{\mathrm{Mg\emph{b}\times (Fe5270+Fe5335)/2}}$}. Galaxies are plotted on top of a model grid which represents solar-\afe\ single-stellar-population (SSP) models of \cite{2003MNRAS.339..897T}. Colour represents \tc. The figure shows that \emph{high-\tc\ ellipticals tend to be younger than more relaxed objects}. In particular, there are no old galaxies with \tc$>$0.13. The galaxy with very large central \hb\ is NGC 2865, a known gas-rich merger remnant.

We derive best-fitting SSP-equivalent age and metallicity (\tssp\ and \zssp) by comparing line-strength indices to predictions from \cite{2003MNRAS.339..897T} models. The comparison is performed on the [MgFe]-H$\beta$ plane shown in Fig.\ref{fig2}. We have verified that  \tssp, the most relevant parameter for our study, is in agreement with values taken from \cite{2005ApJ...621..673T}, \cite{2007A&A...463..455A}, \cite{2007MNRAS.377..759S} and \cite{2008A&A...483...57S} despite the fact that we neglect the effect of \afe. We list \tssp\ values in Table \ref{tab1}.

   \begin{figure}
   \centering
   \includegraphics[width=9cm]{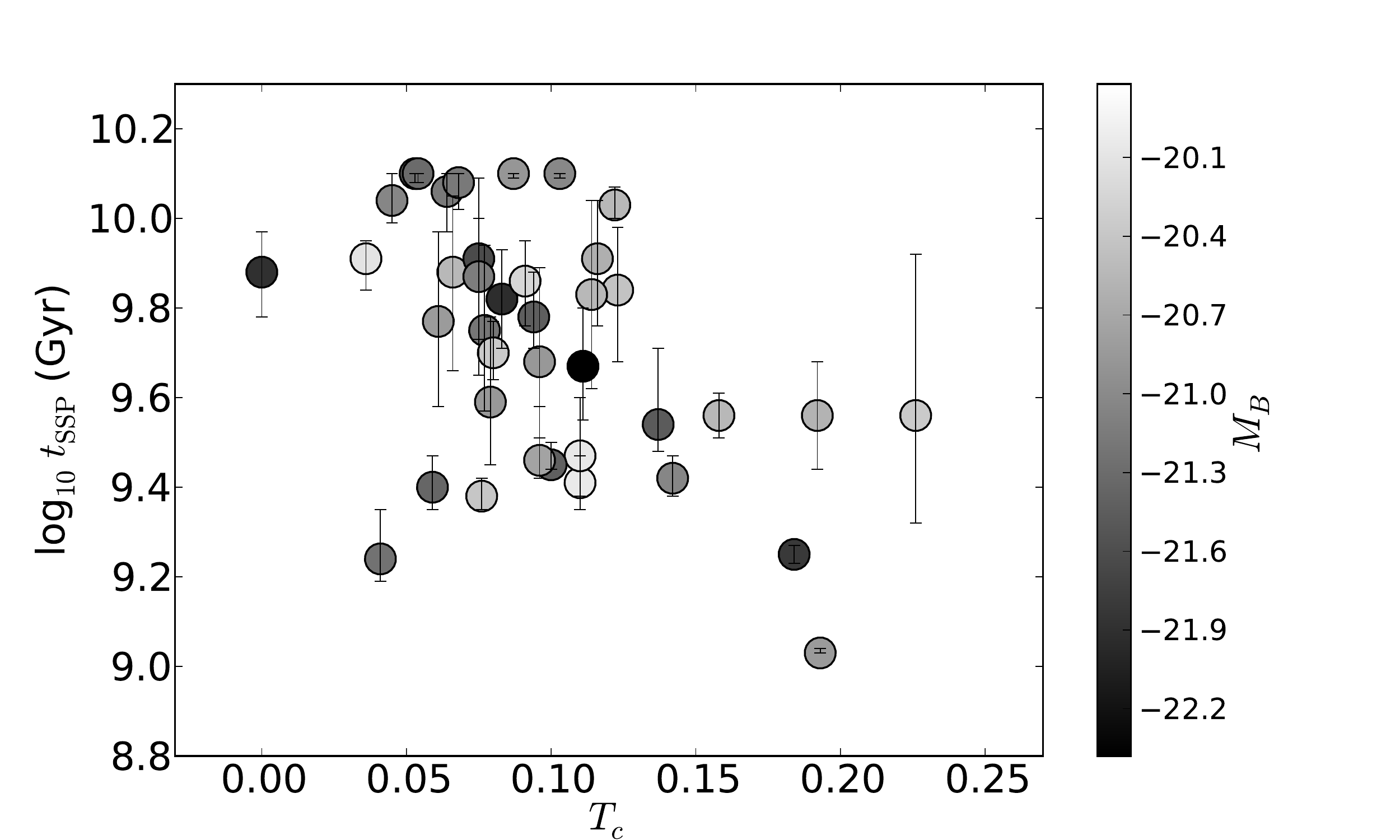}
      \caption{\tssp\ vs. \tc\ for the 38 galaxies in the T09 sample with available line-strength indices. The grey-scale codes $M_B$.}
         \label{fig3}
   \end{figure}

In Fig.\ref{fig3} we plot \tssp\ against \tc, with grey-scale representing $M_B$. While we do not see a clean correlation between \tssp\ and \tc, highly-distrubed ellipticals clearly stand out as they are all fairly young (\tssp$<$4 Gyr). On the contrary, most galaxies with low \tc\ are older, although there are a number of weakly-disturbed objects with young ages too. We note that high-\tc\ galaxies are not younger because of being systematically fainter than more relaxed objects.

If a galaxy hosts stars with a spread in age, \tssp\ is strongly biased towards the age of the youngest stars (see \citealt{2007MNRAS.374..769S} for a thorough discussion of this effect). Therefore, it is likely that galaxies with a \tssp\ of just a few Gyr host recently-formed stars on top of an old, dominant population. We estimate the young-to-total stellar mass fraction \my\  by comparing \hb\ and \mgfe\ to predictions from solar-\afe, two-SSP models built based on \cite{2003MNRAS.339..897T} SSPs. Since the number of free parameters is large, we assume that the old population has an age of 13 Gyr and that the young population has solar metallicity. We let the metallicity of the old population \citep[which dominates \mgfe\ - see][]{2007MNRAS.374..769S} and \my\ free to vary, and fix the age of the young population \ty\ to 300 Myr and 1 Gyr in two different fitting iterations. Given the known degeneracy between \ty\ and \my\ this approach allows us to more easily estimate the allowed range for \my.

We find that ellipticals with \tssp$<$4 Gyr contain 0.5-5\% of their stellar mass in stars formed 300 Myr to $\sim$1 Gyr ago. Given the many approximations involved in this analysis these numbers should be taken with some care. Nevertheless, it is hard to question that many ellipticals, and in particular all highly-disturbed objects, are consistent with having formed up to a few percent of their stellar mass within the past $\sim$1 Gyr. We note that it would have been impossible to reach this conclusion based on optical photometry only. Indeed, the best-fitting two-SSP models predict $B-V$ colours in the range 0.9-1.1 with typical 1$\sigma$ uncertainty below 0.1 dex. The only exception is NGC 2865, which is predicted to have $B-V\sim$0.8.

\section{Discussion and conclusions}

Neutral hydrogen observations and line-strength index analysis of the stellar populations of ellipticals in the T09 sample reveal that a large fraction of these galaxies continued their assembly during the past few Gyr in the presence of a mass of cold gas of \emph{at least} a few percent of the total galaxy stellar mass. This gas is now detected in the form of \hi\ and through young stellar sub-populations.

\hi\ interferometry  reveals that in nearly all cases the detected gas is morphologically/kinematically disturbed and mostly found in tails/filaments, sometimes clearly associated to the optical morphological disturbances. This, and the presence of young stars, confirm T09 conclusion that the assembly of  elliptical galaxies continues to $z$=0. However, it does not agree with their conclusion that  this assembly is essentially gas-free. In fact, \hi\ observations could be regarded as yet one more tool to reveal the on-going assembly of these systems, complementary to deep optical imaging \citep[see][]{2006MNRAS.371..157M,2007A&A...465..787O,2009A&A...498..407G}.

Our estimate of the mass of cold gas recently accreted by ellipticals is in qualitative agreement with the mass accretion rate derived by T09 themselves. They estimate that nearby ellipticals are still growing at $z$=0 at a rate of 20\%-in-mass per Gyr. Evidence of this on-going assembly is particularly strong in the field and in galaxy groups, where most small galaxies around a given elliptical are gas rich. It is therefore reasonable to expect that such a conspicuous mass assembly is accompanied by an accretion of cold gas of \emph{at least} a few percent of the total galaxy stellar mass.

Despite this high incidence of disturbed \hi\ systems, elliptical galaxies with very regular gas distributions do exist \citep[e.g., NGC~2974; more examples can be found in ][]{2006MNRAS.371..157M,2007A&A...465..787O}. These systems could be evolved versions of the galaxies that show disturbed \hi\ at $z$=0, i.e., galaxies for which gas accretion was important many Gyr ago and whose recent history has been quiet enough for this gas to settle.

Another interesting point is that while we find that optically disturbed ellipticals are systematically younger than relaxed objects, and interpret this as evidence of their continuing, gas-rich assembly, not all ellipticals with \hi\ are young (e.g., NGC~5846; other cases are known outside the T09 sample, e.g., NGC~4278 in \citealt{2006MNRAS.371..157M}). Clearly, the dynamics of the accretion plays a very important role in determining the effect of gas on structure and (distribution of) stellar population of the accreting galaxy. Understanding where the young stars and cold gas are, and what their kinematics is, is as important as estimating their mass.

Our results agree with previous work on this subject. \cite{2009arXiv0908.2548S} analyse the stellar population of high-$z$ early-type galaxies selected by \cite{2005AJ....130.2647V} to be ``dry''-merger remnants. They find that highly disturbed objects are on average younger than relaxed ones, and that their line-strength indices are consistent with the presence of a few-percent-in-mass of young stars. This is in agreement with early results from \cite{1990ApJ...364L..33S} who find the \hb\ index to be systematically higher for ellipticals with morphological fine structure at any given absolute magnitude. Finally, \cite{2007AJ....134.1118D} look at the \hi\ properties of a sample of local galaxies selected in the same way as the \cite{2005AJ....130.2647V} objects. They find that many of these systems host a significant mass of \hi\ whose disturbed morphology and kinematics make them consistent with having recently accreted cold gas.

Following the conclusions of a large number of numerical studies, the direct or indirect detection of gas present during the assembly of elliptical galaxies may have important implications for many structural properties of these objects. We have shown that using only observables such as colours measured over large apertures one risks to neglect the evidence of this gas because of the lower sensitivity to the presence of a modest, but nevertheless significant, mass of cold gas or young-stars.

\bibliographystyle{mn2e}
\bibliography{serra}

\begin{table*}
 \centering
% \begin{minipage}{140mm}
  \caption{Sample properties.}
\begin{tabular}{rlrrrrrrrrrrr}
  \hline
\multicolumn{2}{c}{galaxy} & \multicolumn{1}{c}{$d$} & \multicolumn{1}{c}{$M_B$} & \multicolumn{1}{c}{$B-V$} & \multicolumn{1}{c}{\tc} & \multicolumn{1}{c}{\mhi}                     & \multicolumn{1}{c}{H$\beta$} & \multicolumn{1}{c}{Mg$b$} & \multicolumn{1}{c}{$<$Fe$>$} & \multicolumn{1}{c}{\tssp} & \multicolumn{1}{c}{env} & \multicolumn{1}{c}{references} \\ 
\multicolumn{2}{c}{}             & \multicolumn{1}{c}{(Mpc)} &           &             &       & \multicolumn{1}{c}{(10$^8$ \msun)} & \multicolumn{1}{c}{($\AA$)}   & \multicolumn{1}{c}{($\AA$)} &\multicolumn{1}{c}{($\AA$)}       & \multicolumn{1}{c}{(Gyr)} &        &  \\ 
\multicolumn{2}{c}{}             & \multicolumn{1}{c}{(1)}             & \multicolumn{1}{c}{(2)} & \multicolumn{1}{c}{(3)}      &  \multicolumn{1}{c}{(4)}     & \multicolumn{1}{c}{(5)} & \multicolumn{1}{c}{(6)}                         & \multicolumn{1}{c}{(7)}           & \multicolumn{1}{c}{(8)}          &\multicolumn{1}{c}{(9)}                & \multicolumn{1}{c}{(10)}  &\multicolumn{1}{c}{}\\ 
\hline                    

 IC &  1459 & 29.2 & -21.47 & 0.96 & 0.137 &     2.5 & 1.70$\pm$0.10 & 5.38$\pm$0.10 & 3.10$\pm$0.08 &  3.5$^{+1.7} _{-0.4}$ & G &  a,d,v \\ 
 IC &  3370 & 26.8 & -20.59 & 0.89 & 0.192 &  $<$4.1 & 1.97$\pm$0.10 & 4.27$\pm$0.10 & 2.73$\pm$0.09 &  3.6$^{+1.2} _{-0.9}$ & U &  a,e,v \\ 
 IC &  4797 & 28.1 & -20.36 & 0.92 & 0.226 &  $<$4.5 & 1.92$\pm$0.26 & 4.52$\pm$0.18 & 2.75$\pm$0.10 &  3.6$^{+4.7} _{-1.5}$ & G &  a,e,x \\ 
 IC &  4889 & 29.2 & -20.54 & 0.88 & 0.158 &    14.3 & 1.89$\pm$0.04 & 4.18$\pm$0.04 & 3.14$\pm$0.05 &  3.6$^{+0.4} _{-0.4}$ & U &  a,f,y \\ 
NGC &  0584 & 20.1 & -20.40 & 0.92 & 0.076 &     1.4 & 2.08$\pm$0.05 & 4.33$\pm$0.04 & 2.90$\pm$0.03 &  2.4$^{+0.2} _{-0.2}$ & F &  a,g,x \\ 
NGC &  0596 & 21.8 & -20.03 & 0.90 & 0.110 &  $<$0.2 & 2.12$\pm$0.05 & 3.95$\pm$0.04 & 2.81$\pm$0.03 &  2.6$^{+0.4} _{-0.3}$ & F &  a,h,x \\ 
NGC &  0720 & 24.1 & -20.86 & 0.96 & 0.079 &  $<$2.5 & 1.77$\pm$0.12 & 5.17$\pm$0.11 & 2.87$\pm$0.09 &  3.9$^{+2.1} _{-1.1}$ & F &  a,i,x \\ 
NGC &  1199 & 33.1 & -20.49 & 0.97 & 0.067 &  $<$2.5 &             - &             - &             - &                     - & G &    a,j \\ 
NGC &  1209 & 35.8 & -20.62 & 0.95 & 0.116 &  $<$7.3 & 1.54$\pm$0.09 & 4.88$\pm$0.09 & 3.17$\pm$0.08 &  8.1$^{+2.8} _{-2.4}$ & G &  a,e,v \\ 
NGC &  1395 & 24.1 & -21.43 & 0.94 & 0.094 &  $<$3.3 & 1.62$\pm$0.05 & 5.21$\pm$0.04 & 2.93$\pm$0.03 &  6.0$^{+1.6} _{-0.9}$ & F &  a,e,x \\ 
NGC &  1399 & 20.0 & -21.16 & 0.98 & 0.064 &  $<$2.3 & 1.43$\pm$0.08 & 5.20$\pm$0.16 & 2.95$\pm$0.10 & 11.5$^{+1.1} _{-2.1}$ & C &  a,e,w \\ 
NGC &  1407 & 28.8 & -21.92 & 0.93 & 0.083 &  $<$4.7 & 1.67$\pm$0.07 & 4.88$\pm$0.06 & 2.85$\pm$0.03 &  6.6$^{+1.9} _{-1.5}$ & G &  a,e,x \\ 
NGC &  2865 & 37.8 & -20.84 & 0.78 & 0.193 &     5.3 & 3.19$\pm$0.05 & 3.06$\pm$0.04 & 2.69$\pm$0.03 &  1.1$^{+0.0} _{-0.0}$ & F &  a,k,z \\ 
NGC &  2974 & 21.5 & -20.05 & 0.95 & 0.110 &     5.8 & 1.95$\pm$0.10 & 4.77$\pm$0.11 & 2.78$\pm$0.10 &  3.0$^{+1.0} _{-0.6}$ & F &  a,l,v \\ 
NGC &  2986 & 30.7 & -21.03 & 0.99 & 0.045 &  $<$5.3 & 1.48$\pm$0.06 & 4.97$\pm$0.05 & 2.92$\pm$0.03 & 11.0$^{+1.6} _{-1.2}$ & F &  b,e,x \\ 
NGC &  3078 & 35.2 & -21.01 & 0.97 & 0.103 &  $<$7.0 & 1.12$\pm$0.09 & 5.20$\pm$0.07 & 3.16$\pm$0.04 & 12.6$^{+0.0} _{-0.3}$ & F &  a,e,x \\ 
NGC &  3258 & 32.1 & -20.42 & 0.92 & 0.123 &  $<$5.8 & 1.57$\pm$0.09 & 5.08$\pm$0.09 & 3.11$\pm$0.09 &  6.9$^{+2.6} _{-2.1}$ & G &  a,e,v \\ 
NGC &  3268 & 34.8 & -20.87 & 0.96 & 0.087 &  $<$6.9 & 1.01$\pm$0.11 & 4.97$\pm$0.11 & 2.64$\pm$0.10 & 12.6$^{+0.0} _{-0.3}$ & C &  a,e,v \\ 
NGC &  3557 & 45.7 & -22.38 & 0.87 & 0.111 & $<$11.8 & 1.75$\pm$0.08 & 4.72$\pm$0.08 & 3.00$\pm$0.08 &  4.7$^{+1.6} _{-1.1}$ & G &  a,e,v \\ 
NGC & 3557B & 38.2 & -19.82 & 0.86 & 0.182 &  $<$8.3 &             - &             - &             - &                     - & G &    c,e \\ 
NGC &  3585 & 20.0 & -20.98 & 0.91 & 0.048 &  $<$2.3 &             - &             - &             - &                     - & F &    a,e \\ 
NGC &  3640 & 27.0 & -21.03 & 0.92 & 0.142 &  $<$0.2 & 1.96$\pm$0.06 & 4.57$\pm$0.08 & 3.09$\pm$0.09 &  2.6$^{+0.3} _{-0.2}$ & F &  a,i,y \\ 
NGC &  3706 & 37.4 & -20.96 & 0.93 & 0.120 &  $<$7.9 &             - &             - &             - &                     - & U &    c,e \\ 
NGC &  3904 & 28.3 & -20.80 & 0.94 & 0.108 &    12.7 &             - &             - &             - &                     - & U &    a,m \\ 
NGC &  3923 & 22.9 & -21.41 & 0.95 & 0.100 &  $<$3.0 & 1.87$\pm$0.08 & 5.12$\pm$0.07 & 3.07$\pm$0.04 &  2.8$^{+0.3} _{-0.1}$ & U &  a,e,x \\ 
NGC &  3962 & 35.3 & -21.35 & 0.95 & 0.059 &    27.9 & 2.02$\pm$0.08 & 4.79$\pm$0.07 & 2.81$\pm$0.07 &  2.5$^{+0.4} _{-0.3}$ & F &  a,n,v \\ 
NGC &  4105 & 26.5 & -20.85 & 0.87 & 0.109 &     7.8 &             - &             - &             - &                     - & F &    a,o \\ 
NGC &  4261 & 31.6 & -21.26 & 0.98 & 0.053 &  $<$5.7 & 1.34$\pm$0.06 & 5.11$\pm$0.04 & 3.01$\pm$0.04 & 12.6$^{+0.0} _{-0.6}$ & C &  a,e,x \\ 
NGC &  4365 & 20.4 & -21.16 & 0.97 & 0.070 &  $<$0.4 &             - &             - &             - &                     - & C &    a,i \\ 
NGC &  4472 & 16.3 & -21.90 & 0.97 & 0.000 &     0.5 & 1.62$\pm$0.06 & 4.85$\pm$0.06 & 2.91$\pm$0.05 &  7.6$^{+1.7} _{-1.6}$ & C &  a,p,x \\ 
NGC &  4636 & 14.7 & -20.54 & 0.93 & 0.066 &     8.1 & 1.58$\pm$0.13 & 5.07$\pm$0.12 & 2.96$\pm$0.16 &  7.6$^{+3.6} _{-3.0}$ & F &  a,q,v \\ 
NGC &  4645 & 29.9 & -20.12 & 0.95 & 0.000 &  $<$5.1 &             - &             - &             - &                     - & C &    a,e \\ 
NGC &  4696 & 35.5 & -21.62 & 0.94 & 0.075 &  $<$7.1 & 1.75$\pm$0.19 & 4.52$\pm$0.25 & 2.34$\pm$0.26 &  8.1$^{+4.2} _{-3.7}$ & C &  a,e,v \\ 
NGC &  4697 & 11.7 & -20.24 & 0.92 & 0.091 &  $<$0.6 & 1.75$\pm$0.07 & 4.08$\pm$0.05 & 2.77$\pm$0.04 &  7.2$^{+1.7} _{-1.5}$ & C &  a,i,x \\ 
NGC &  4767 & 32.8 & -20.57 & 0.93 & 0.000 &  $<$6.1 &             - &             - &             - &                     - & C &    a,e \\ 
NGC &  5011 & 41.9 & -21.20 & 0.89 & 0.077 &  $<$9.9 & 1.78$\pm$0.13 & 4.61$\pm$0.15 & 2.70$\pm$0.15 &  5.6$^{+3.1} _{-1.9}$ & U &  a,e,v \\ 
NGC &  5018 & 40.5 & -21.80 & 0.85 & 0.184 &     7.4 & 2.39$\pm$0.06 & 3.51$\pm$0.09 & 2.85$\pm$0.10 &  1.8$^{+0.1} _{-0.1}$ & F &  c,r,y \\ 
NGC &  5044 & 31.2 & -21.23 & 0.98 & 0.041 &  $<$5.5 & 2.44$\pm$0.28 & 4.58$\pm$0.25 & 2.56$\pm$0.24 &  1.7$^{+0.5} _{-0.2}$ & G &  a,e,v \\ 
NGC &  5061 & 29.1 & -21.44 & 0.85 & 0.104 &  $<$4.8 &             - &             - &             - &                     - & F &    a,e \\ 
NGC &  5077 & 38.0 & -20.82 & 0.98 & 0.061 &    15.4 & 1.66$\pm$0.12 & 4.92$\pm$0.11 & 2.98$\pm$0.11 &  5.9$^{+3.4} _{-2.1}$ & G &  c,d,v \\ 
NGC &  5576 & 25.5 & -20.40 & 0.88 & 0.122 &  $<$0.4 &             - &             - &             - &                     - & F &    a,s \\ 
NGC &  5638 & 26.3 & -20.11 & 0.94 & 0.036 &  $<$0.3 & 1.65$\pm$0.04 & 4.64$\pm$0.04 & 2.84$\pm$0.04 &  8.1$^{+0.8} _{-1.2}$ & F &  a,t,x \\ 
NGC &  5812 & 26.9 & -20.36 & 0.94 & 0.080 &  $<$4.1 & 1.70$\pm$0.04 & 4.81$\pm$0.04 & 3.06$\pm$0.04 &  5.0$^{+0.9} _{-0.6}$ & F &  a,e,x \\ 
NGC &  5813 & 32.2 & -21.30 & 0.95 & 0.054 &  $<$1.0 & 1.42$\pm$0.07 & 4.65$\pm$0.05 & 2.67$\pm$0.05 & 12.6$^{+0.0} _{-0.6}$ & G &  a,i,x \\ 
NGC &  5846 & 24.9 & -21.16 & 0.98 & 0.068 &     3.5 & 1.45$\pm$0.07 & 4.93$\pm$0.05 & 2.86$\pm$0.04 & 12.0$^{+0.6} _{-1.6}$ & G &  a,o,x \\ 
NGC &  5898 & 29.1 & -20.54 & 0.92 & 0.114 &  $<$4.8 & 1.78$\pm$0.17 & 4.30$\pm$0.17 & 2.56$\pm$0.15 &  6.8$^{+4.2} _{-2.6}$ & G &  a,u,v \\ 
NGC &  5903 & 33.9 & -21.13 & 0.89 & 0.075 &     9.2 & 1.68$\pm$0.10 & 4.44$\pm$0.08 & 2.90$\pm$0.05 &  7.4$^{+2.6} _{-2.0}$ & G &  a,u,x \\ 
NGC &  6861 & 28.1 & -20.44 & 0.95 & 0.123 &  $<$4.5 &             - &             - &             - &                     - & G &    a,e \\ 
NGC &  6868 & 26.8 & -20.75 & 0.97 & 0.096 &  $<$4.1 & 1.84$\pm$0.10 & 5.05$\pm$0.10 & 3.04$\pm$0.09 &  2.9$^{+0.9} _{-0.3}$ & G &  a,e,v \\ 
NGC &  6958 & 33.1 & -20.53 & 0.86 & 0.122 &  $<$6.2 & 1.63$\pm$0.05 & 3.89$\pm$0.04 & 2.75$\pm$0.03 & 10.7$^{+1.0} _{-0.7}$ & C &  c,e,x \\ 
NGC &  7029 & 38.4 & -20.72 & 0.86 & 0.085 &  $<$8.4 &             - &             - &             - &                     - & G &    a,e \\ 
NGC &  7144 & 24.5 & -20.38 & 0.91 & 0.100 &  $<$3.4 &             - &             - &             - &                     - & G &    a,e \\ 
NGC &  7192 & 37.8 & -20.85 & 0.92 & 0.096 &  $<$8.1 & 1.71$\pm$0.12 & 4.66$\pm$0.11 & 3.17$\pm$0.10 &  4.8$^{+3.0} _{-1.6}$ & F &  a,e,v \\ 
NGC &  7196 & 45.1 & -21.03 & 0.91 & 0.171 & $<$11.5 &             - &             - &             - &                     - & G &    a,e \\ 
NGC &  7507 & 25.0 & -20.85 & 0.94 & 0.084 &  $<$3.5 &             - &             - &             - &                     - & F &    a,e \\ 
\hline
\end{tabular}

(1) Distance from: ($a$) \cite{2001ApJ...546..681T}; ($b$) NED recessional velocity corrected for Virgo infall ($h=0.73$); ($c$) NED $z$-independent. (2) $M_B$ derived from HyperLeda $btc$. (3) Colour as in T09. (4) \tc\ from T09. (5) \mhi\ adopting the total \hi\ flux from: ($d$) Oosterloo (priv.~comm.); ($e$) HIPASS spectra \citep{2004MNRAS.350.1195M}; ($f$) \cite{2007A&A...465..787O}; ($g$) \cite{1990AJ.....99.1740H}; ($h$) \cite{2000AJ....120.1946S}; ($i$) \cite{1985AJ.....90..454K}; ($j$) \cite{1994A&A...286..389H}; ($k$)\cite{1995ApJ...444L..77S}; ($l$) \cite{2008MNRAS.383.1343W}; ($m$) \cite{1977A&A....60L..23B}; ($n$) \cite{1979A&A....76..176B}; ($o$) \cite{1979A&A....74..172B}; ($p$) \cite{1994AJ....108..844M}; ($q$) \cite{1978AJ.....83...11K}; ($r$) \cite{1988ApJ...330..684K}; ($s$) \cite{1984ApJ...280..107L}; ($t$) \cite{1987ApJS...63..515L}; ($u$) \cite{1990ApJ...357..426A}; see text for the calculation of the upper limits. (6-8) Line-strength indices from: ($v$) \cite{2006A&A...445...79A}; ($w$) Trager (priv.~comm.); ($x$) \cite{2005ApJ...621..673T}; ($y$) \cite{2008A&A...483...57S}; ($z$) \cite{2007MNRAS.377..759S}; $<$Fe$>$=(Fe5270+Fe5335)/2. (9) \tssp\ derived in this work. (10) Environment from T09.
%\end{minipage}
\label{tab1}
\end{table*}

\label{lastpage}

\end{document}